\documentclass[preprint,12pt]{elsarticle}
\usepackage{graphicx}
\usepackage{dcolumn}
\usepackage{bm}
\usepackage[hidelinks]{hyperref}
\usepackage{academicons}
\usepackage{xcolor}
\usepackage{xcolor}
\hypersetup{
    colorlinks,
    linkcolor={blue},
    citecolor={red},
    urlcolor={blue!80!black}
}
\usepackage[framemethod=TikZ]{mdframed}
\usepackage[most]{tcolorbox}
\usepackage{longtable}
\usepackage{mathcomp}
\usepackage{pifont}
\usepackage{xcolor}
\usepackage{amsthm}
\usepackage{PGF}
\usepackage{pgfpages}
\usepackage{amsmath}
\usepackage{amsfonts}
\usepackage{amssymb}
\usepackage{graphicx}
\usepackage{bm}
\usepackage{hyperref}
\usepackage{lipsum}
\usepackage{array}
\usepackage{booktabs}
\usepackage{tikz}
\usepackage{titlesec}
\usepackage{hyperref}
\hypersetup{bookmarksnumbered}
\def\be{\begin{equation}}
\def\ee{\end{equation}}

\usepackage{amsmath}
\usepackage{empheq}
\usepackage[most]{tcolorbox}
\newtcbox{\mymath}[1][]{%
    nobeforeafter, math upper, tcbox raise base,
    enhanced, colframe=blue!30!black,
    colback=blue!30, boxrule=1pt,
    #1}
\usepackage{mdframed}
\usepackage{lipsum}
\usepackage{PGF}
\usepackage{pgfkeys}
\usepackage{pgffor}
\usepackage{pgfcalendar}
\usepackage{pgfpages}
\usepackage{tikz}
\usetikzlibrary{calc}
\usepackage{array}
\usepackage{booktabs}
\usepackage{multirow}
\usepackage{latexsym}
\usepackage[english]{babel}
\usepackage[most]{tcolorbox}

\newtcolorbox{myquote}[1][]{%
    colback=black!5,
    colframe=black!5,
    notitle,
    sharp corners,
    borderline west={2pt}{0pt}{red!80!black},
    enhanced,
    breakable,
    }
\usepackage{enumitem}

{\begin{itemize}\item[]}
{\end{itemize}}
\usepackage{graphicx}
\usepackage{mathtools}

\usepackage{amssymb}
\begin{document}

\begin{frontmatter}


\title{\textcolor[rgb]{0.00,0.40,0.80}{\bfseries{\boldmath
Inflation and Isotropization in Quintom Cosmology
}}}

\author[inst1]{Behzad Tajahmad}
\affiliation[inst1]{behzadtajahmad@yahoo.com}


\begin{abstract}
This paper studies inflation and isotropization in the quintom model in the Bianchi I, Bianchi III, and Kantowski-Sachs backgrounds. First, we investigate inherent properties and generalize Heusler's proposition. Then by the use of the dynamical system approach, we consider the system in multiplicative and collective modes of potentials. The conclusions of Collins and Hawking and also Burd and Barrow are discussed.
\end{abstract}


\end{frontmatter}

\section{Introduction\label{sec:intro}}
As a solution to homogeneity, isotropy, and the horizon problem, inflation was introduced in the standard cosmological model by Guth~\cite{1}.
The latter is well explained due to the fact that inflation is characterized by a power-law or exponential expansion of the universe and at the same time a quasi-constant behavior of the Hubble horizon.

However, since the Friedman-Lemaitre metric is used from the start, homogeneity and isotropy have not been well explained. The real solution to the problem is to start with an arbitrary metric, show that inflation occurs, and then show that the universe evolves towards a Friedman-Lemaitre metric. Numerical simulations were conducted on cosmologies with spherical inhomogeneous configurations to determine the onset of inflation~\cite{3,2}. Using the long wavelength iteration scheme, a semi-numerical analysis was performed on inhomogeneous, quasi-isotropic universes~\cite{6,5,4}.
In their study, the researchers demonstrated that a large level of initial inhomogeneity suppresses inflation. Due to the difficulty of the task, one can try to solve the isotropy problem by applying a homogeneous but anisotropic metric as a first step. This approach was first adopted by Collins and Hawking in~\cite{7}.
They indicated that the isotropy problem can only be solved for the Bianchi types $\mathrm{I}$, $\mathrm{V}$, $\mathrm{VII_{O}}$, and $\mathrm{VII_{h}}$ filled by matter satisfying the dominant energy condition.
They also showed that the set of spatially homogeneous cosmological models that approach isotropy at infinite times is of measure zero in the space of all spatially homogeneous models.

There was hope that the cosmic no-hair theorem could be derived with an inflationary stage, where the dominant energy condition is violated.
In a Bianchi-type universe, for real scalar fields with a convex positive potential and a vanishing local minimum, Heusler demonstrated that there is no no-hair theorem~\cite{8}.
As a matter of fact, to approach isotropy, a Friedman-Lemaitre model must be compatible with the Lie group underlying the Bianchi-type metric.

In addition to Bianchi-type metrics, the Kantowski-Sachs model offers descriptions of spatially homogeneous universes as well.
Several authors have studied the model with or without a cosmological constant and a perfect fluid description of matter~\cite{13,11,9}.
An anisotropic asymptotical behavior of the model was found in their study.

Quintom model is one of the dynamical dark energy models~\cite{14,15,16,17}. However, it was proposed as a solution to the accelerated expansion problem but such models should be examined in different stages of universe evolution to answer the question of whether we can reach a unified model or not. Quintom and in general multi-scalar field models have been widely studied in the literature. See for instance refs.~\cite{er1,er2,er3,er4,er5,er6} and references therein.\\
In this work, we study in detail a quintom scalar field with convex positive potentials, but not necessarily with local minimums, in a Bianchi I, a Bianchi III, and a Kantowski-Sachs models. We want to find out under what conditions inflation and/or isotropy may occur. It should be mentioned that for the phantom scalar field, this task has been performed in refs.~\cite{prd98,BBa}.

\section{The model and equations\label{sec:intro}}
Let us start with the gravitational action of the form~\cite{action}
\begin{align}
S=\int d^{4}x \sqrt{-g} \biggl[- \frac{R}{16 \pi G}+
\frac{1}{2}\partial_{\mu}\varphi_{1}\partial^{\mu}\varphi_{1}
-\frac{1}{2}\partial_{\mu}\varphi_{2}\partial^{\mu}\varphi_{2}
+V(\varphi_{1},\varphi_{2}) \biggl]
\end{align}
where $g$ is the determinant of metric, $R$ is the Ricci scalar, and $\varphi_{1}$ and $\varphi_{2}$  are real scalar fields having potential $V(\varphi_{1},\varphi_{2})$. We consider the action in a homogeneous universe of the form
\begin{align}
ds^{2}=-dt^{2}+a(t)^{2}dr^{2}+b(t)^{2} \left( d\theta^{2}+f(k,\theta)^{2}
d\phi^{2} \right),
\end{align}
where
\begin{align*}
f(k,\theta)=\left\{
       \begin{array}{lll}
      \theta, &\text{for }k=0:& \hbox{Bianchi type I (B-I);} \\
      \sinh \theta, &\text{for }k=-1:& \hbox{Bianchi type III (B-III);} \\
      \sin \theta, &\text{for }k=1:& \hbox{Kantoweski-Sachs (KS).}
       \end{array}
     \right.
\end{align*}
By varying the action with respect to the metric we get the corresponding Einstein field equations:
\begin{align}
\label{feq1}
&2H_{a}H_{b}+H_{b}^{2}+\frac{k}{b^{2}}=8\pi G
\biggl(\frac{1}{2}\dot{\varphi}_{1}^{2}-\frac{1}{2}\dot{\varphi}_{2}^{2}+V \biggl),\\
\label{feq2}
&2\dot{H}_{b}+3H_{b}^{2}+\frac{k}{b^{2}}=8\pi G\biggl(-\frac{1}{2}\dot{\varphi}_{1}^{2}+\frac{1}{2}\dot{\varphi}_{2}^{2}+V \biggl),\\
\label{feq3}
&\dot{H}_{a}+\dot{H}_{b}+H_{a}^{2}+H_{b}^{2}+H_{a}H_{b}=8\pi G\biggl(-\frac{1}{2}\dot{\varphi}_{1}^{2}+\frac{1}{2}\dot{\varphi}_{2}^{2}+V \biggl),
\end{align}
in which the dot denotes a differentiation with respect to time $t$ and the directional Hubble parameters are defined as $H_{a}=\dot{a}/a$ and $H_{b}=\dot{b}/b$.\\
The Klein-Gordon equations for the scalar fields $\varphi_{1}$ and $\varphi_{2}$ are obtained by varying the action with respect to $\varphi_{1}$ and $\varphi_{2}$, respectively:
\begin{align}
\label{feq4}
\ddot{\varphi}_{1}+(H_{a}+2H_{b})\dot{\varphi}_{1}+\frac{\partial V}{\partial \varphi_{1}}&=0,\\
\label{feq5}
\ddot{\varphi}_{2}+(H_{a}+2H_{b})\dot{\varphi}_{2}-\frac{\partial V}{\partial \varphi_{2}}&=0.
\end{align}
Planck mass will be used to express all quantities in the following. Hence we set $8\pi G=1$.\\
As a consequence of Bianchi identities, the system is fully determined by the independent equations (\ref{feq1})-(\ref{feq2}) and (\ref{feq4})-(\ref{feq5}), because eq.~(\ref{feq3}) follows from the others.\\
Utilizing the shear scalar $\sigma = \sqrt{1/3}(H_{a} - H_{b})$ and the expansion scalar $\Theta =H_{a}+2H_{b}$, this system of equations can be written as a set of six first-order differential equations, which are $k$ independent, and a constraint that is conserved in the evolution:
\begin{align}
\label{pdeq1}&\dot{\varphi}_{1}=\psi_{1},\\
\label{pdeq2}&\dot{\varphi}_{2}=\psi_{2},\\
\label{pdeq3}&\dot{\Theta}=\frac{-1}{3}\Theta^{2}-2\sigma^{2}
-\psi_{1}^{2}+\psi_{2}^{2}+V,\\
\label{pdeq4}&\dot{\sigma}=\frac{-1}{3\sqrt{3}}\Theta^{2}
+\frac{1}{\sqrt{3}}\sigma^{2}-\Theta \sigma +\frac{1}{\sqrt{3}}
\biggl( \frac{1}{2}\psi_{1}^{2}-\frac{1}{2}\psi_{2}^{2}+V \biggl)\\
\label{pdeq5}&\dot{\psi}_{1}=-\Theta \psi_{1}-\frac{\partial V}{\partial \varphi_{1}},\\
\label{pdeq6}&\dot{\psi}_{2}=-\Theta \psi_{2}+\frac{\partial V}{\partial \varphi_{2}},\\
\label{pdeq7}&\frac{1}{3}\Theta^{2}+\frac{k}{b^{2}}=\sigma^{2}
+\frac{1}{2}\psi_{1}^{2}-\frac{1}{2}\psi_{2}^{2}+V,
\end{align}
where the last equation is the constraint equation.\\
The selection of the elements of this system $\{\psi_{1}, \psi_{2}, \Theta, \sigma\}$ is completely meaningful: $\psi_{1}$ and $\psi_{2}$ are selected just for reducing the order of equations, $\Theta$ identifies the expansion/contraction of the universe which is positive for an expanding universe, and $\sigma$ measures the anisotropy. As a criterium of isotropization, rather than using the vanishing $\sigma$, we will utilize a stronger condition~\cite{7}: $(\sigma / \Theta) \to 0$ as $t \to \infty$.\\
Due to the absence of $k$ in eqs.~(\ref{pdeq1})-(\ref{pdeq6}), the solutions of them do not depend on the type of the homogeneous model. Because of the conservation of eq.~(\ref{pdeq7}), the homogeneous model only has to be specified at the beginning. In five-dimensional space with the coordinate $\{\Theta, \sigma, \psi_{1},\psi_{2},\sqrt{V}\}$, the constraint equation forms a hypersurface so that the B-I solutions are on it, the B-III ones are in it and the KS solutions remain outside the hypersurface.
\section{Intrinsic properties of dynamical systems}
The following section generalizes one of Heusler's findings (i.e. proposition. 1)~\cite{8}. \\
First of all, in this paper, we restrict ourselves to the expanding universe ($\Theta >0$), and to positive and convex function $V$.
Furthermore, in this section, we suppose that scalar fields are real.\\
For B-I and B-III universes (i.e. $k \leq 0$), the following inequality relation is obtained by the use of eqs.~(\ref{pdeq3}) and (\ref{pdeq7}):
\begin{align}
\label{ieq1}
\dot{\Theta} \leq V-\frac{1}{3}\Theta^{2} \leq 0,
\end{align}
provided that $\psi_{2} \leq \psi_{1}$ or more precisely $\psi_{2}^{2} \leq 2\sigma^{2}+\psi_{1}^{2}$. Therefore, $\Theta$ decreases with time. Utilizing eqs.~(\ref{ieq1}),~(\ref{pdeq3}), and~(\ref{pdeq7}) we arrive at
\begin{align}
\label{ieq2}
\dot{\Theta}+\Theta^{2}=\frac{-2k}{b^{2}}+3V \geq 0,
\end{align}
again under the aforementioned condition.\\
In terms of average scale factor, $a_{\mathrm{ave.}}=\sqrt[3]{ab^{2}}$, we can write
\begin{align*}
\Theta =3\frac{\dot{a}_{\mathrm{ave.}}}{a_{\mathrm{ave.}}}.
\end{align*}
Using eqs.~(\ref{ieq1}) and (\ref{ieq2}) we easily obtain: $\ddot{a}_{\mathrm{ave.}}/a_{\mathrm{ave.}}>0$ meaning that $\dot{a}_{\mathrm{ave.}}$ increases with time and if we set $\dot{a}_{\mathrm{ave.}}|_{t_{0}}>0$ then $\Theta$ will remain positive after $t_{0}$ and so it refers to an expanding universe. Therefore $\Theta$ approaches a positive constant value, say $\Theta_{\infty}$, as the universe ages.\\
By the use of eqs.~(\ref{pdeq3}) and (\ref{pdeq7}) we get:
\begin{align}
\label{ieq3}
\dot{\Theta}=\frac{k}{b^{2}}-3\sigma^{2}-\frac{3}{2}\psi_{1}^{2}
+\frac{3}{2}\psi_{2}^{2}
\end{align}
By combining (\ref{ieq1}) with (\ref{ieq3}) and using the fact that $\dot{\Theta} \to 0$ as $t \to \infty$ it is found that $(k/b^{2})$, $\sigma$, and $\left(\psi_{1}^{2}-\psi_{2}^{2}\right)$ must vanish as $t \to \infty$ and consequently by utilizing (\ref{pdeq7}) we reach $\Theta \to \sqrt{3V}$ as $t \to \infty$. On the other hand, our system must land at the minimum of the potential, say $V_{0}$, because both (\ref{pdeq5}) and (\ref{pdeq6}) describe damped harmonic oscillators, and therefore we have $\Theta \to \Theta _{\infty}=\sqrt{3V_{0}}$ as $t \to \infty$. For exponential potential decaying with $\varphi_{1}$ and $\varphi_{2}$, the only minimum is at infinity and $V_{0}=0$.\\
Now we can generalize proposition 1 in ref.~\cite{8} as follows:\\

\textbf{Theorem 1.}\\
Let $V(\varphi_{1},\varphi_{2})\geq V_{0} \geq 0$ for all $\varphi_{1}$ and $\varphi_{2}$. If there exists a time $t_{0}$ with $\Theta (t_{0}) \geq 0$ then we have the following properties for B-I and B-III universes under the condition $\dot{\varphi}_{2} \leq \dot{\varphi}_{1}$:\\
\begin{enumerate}
	 \item $\Theta (t) \geq 0$, $\dot{\Theta}(t) \leq 0$ for all $t \geq t_{0}$
  \item $\sigma$, $\frac{k}{b^{2}}$, $\left(\dot{\varphi}_{1}^{2}-\dot{\varphi}_{2}^{2}\right)\; \to 0$, and $\Theta \to \sqrt{3V_{0}}$ as $t \to \infty$.
\end{enumerate}
In this theorem, $V_{0}$ is the minimum of the potential. In particular, for an exponentially potential we have $V_{0}=0$.\\
The B-I and B-III converge exponentially to the isotropic De Sitter universe if $V_{0}>0$~\cite{20} while if $V_{0}=0$ at some finite values $\varphi_{10}$ and $\varphi_{20}$, then isotropization occurs just in the B-I case~\cite{8}. There is no definitive answer to the question of isotropization for an exponential potential in either a B-I or a B-III model using the above theorem because $\Theta$ and $\sigma$ vanish together. In this case, to determine the conditions under which isotropy and/or inflation can occur, the asymptotic solution of the dynamical system must be found. This task is performed in the next section in detail.

\section{Investigation of the dynamical system with an exponential potential}
It is crucial to determine the asymptotic behavior near the critical points in order to qualitatively discuss the system of differential equations~(\ref{pdeq1})-(\ref{pdeq6}).
The usual methods of treating the problem do not work effectively:
Linearizing or even transforming to a normal form is unproductive since the singular points are highly nonhyperbolic. The Lyapunov function method also fails to solve the problem because the known candidate namely the constraint equation has no isolated root. On the other hand, guessing another candidate for the Lyapunov function is a hard task. The problems arise from the fact that the right-hand side of equations only contains quadratic terms. For such a system, the standard approach is that we divide all variables but one by the remaining variable and rewrite the system in terms of these new variables. The number of critical points is affected by the choice of this special variable. In our case of study, $\Theta$ is the best variable which leads to the maximum number of critical points (CPs). Other variables like the potential or etc yield less number of CPs. The form of the potential function $V(\varphi_{1},\varphi_{2})$ should be clarified at this stage. Two options seem interesting: \textit{i}. Class-1: Multiplicative mode: $V(\varphi_{1},\varphi_{2})=V_{1}(\varphi_{1})V_{2}(\varphi_{2})=V_{01}\exp (-\lambda_{1}\varphi_{1})V_{02}\exp (-\lambda_{2}\varphi_{2})$; \textit{ii}. Class-2: Collective mode: $V(\varphi_{1},\varphi_{2})=V_{1}(\varphi_{1})+V_{2}(\varphi_{2})=V_{01}\exp (-\lambda_{1}\varphi_{1})+V_{02}\exp (-\lambda_{2}\varphi_{2})$. In what follows we study both cases separately in detail.

\subsection{Class-1: Multiplicative mode: $V(\varphi_{1},\varphi_{2})=V_{1}(\varphi_{1})V_{2}(\varphi_{2})$}

In this subsection, we investigate the potential of the form $V(\varphi_{1},\varphi_{2})=V_{1}(\varphi_{1})V_{2}(\varphi_{2})=V_{01}\exp (-\lambda_{1}\varphi_{1})V_{02}\exp (-\lambda_{2}\varphi_{2})$.\\
Defining the variables $S$, $U$, $P_{1}$ and $P_{2}$ by
\begin{align}
\label{d}
S=\frac{\sigma}{\Theta}, \quad U=\frac{\sqrt{V}}{\Theta}, \quad
P_{1}=\frac{\psi_{1}}{\Theta}, \quad P_{2}=\frac{\psi_{2}}{\Theta},
\end{align}
the eqs.~(\ref{pdeq1})-(\ref{pdeq6}) transform into
\begin{align}
\label{deq1}
\Theta^{\prime}=&\Theta \biggl( -\frac{1}{3}-2S^{2}+U^{2}
-P_{1}^{2}+P_{2}^{2} \biggl),\\
\label{deq2}
S^{\prime}=&\frac{-1}{3\sqrt{3}}-\frac{2}{3}S+\frac{1}{\sqrt{3}}S^{2}
+2S^{3}+P_{1}^{2}\biggl( \frac{1}{2\sqrt{3}}+S \biggl)\nonumber\\&
-P_{2}^{2}\biggl( \frac{1}{2\sqrt{3}}+S \biggl)
+U^{2}\biggl( \frac{1}{\sqrt{3}}-S \biggl),\\
\label{deq3}
U^{\prime}=&U\biggl( \frac{1}{3}+2S^{2}-U^{2}-\frac{\lambda_{1}}{2}
P_{1}-\frac{\lambda_{2}}{2}P_{2}+P_{1}^{2}-P_{2}^{2} \biggl),\\
\label{deq4}
P_{1}^{\prime}=&P_{1}\biggl( -\frac{2}{3}+2S^{2}+P_{1}^{2}-P_{2}^{2} \biggl)
+U^{2}(\lambda_{1}-P_{1}),\\
\label{deq5}
P_{2}^{\prime}=&P_{2}\biggl( -\frac{2}{3}+2S^{2}+P_{1}^{2}-P_{2}^{2} \biggl)
-U^{2}(\lambda_{2}+P_{2}),
\end{align}
and the constraint equation is
\begin{align}
\label{deq6}
S^{2}+U^{2}+\frac{1}{2}P_{1}^{2}-\frac{1}{2}P_{2}^{2}+\frac{k}{b^{2}}
=\frac{1}{3},
\end{align}
where the prime denotes $^{\prime}=d/d\tau =\Theta^{-1}d/\mathrm{d}t$. If we use dot instead of prime, then we will have an extra $\Theta$ on the right-hand side of all equations. Its consequence is just the production of a trivial family of CPs whose $\Theta$ is zero meaning there is no expansion.
In our system, $\Theta$ is given by a simple integration of~(\ref{deq1}).\\
According to eq.~(\ref{deq6}), for the B-I universe ($k = 0$), it describes the surface of a hyper-ellipsoid that separates the KS (inside) from the B-III (outside) solutions. Obviously, for the B-I universe ($k=0$), the problem reduces to a three-dimensional submanifold $W$ in the space of the variables $(S,U,P_{1},P_{2})$, defined by
\begin{align*}
&W=f^{-1}\left( \left\{ \frac{1}{3} \right\} \right)\\&
\text{with } f(S,U,P)=S^{2}+U^{2}+\frac{1}{2}P_{1}^{2}-\frac{1}{2}P_{2}^{2}.
\end{align*}
Equating the left-hand sides of eqs.~(\ref{deq1})-(\ref{deq5}) to zero, five CPs and one family of CPs are found:
\begin{align*}
&\textbf{CP1: }S=\frac{-1}{2\sqrt{3}},\quad U=0,\quad P_{1}=0,\quad P_{2}=0;\\
&\textbf{CP2: }S=0,\quad U=\sqrt{\frac{6+\lambda_{2}^{2}-\lambda_{1}^{2}}{18}},\quad P_{1}=\frac{\lambda_{1}}{3}, \\
&\qquad \; P_{2}=\frac{-\lambda_{2}}{3};\\
&\textbf{CP3: }S=0,\quad U=-\sqrt{\frac{6+\lambda_{2}^{2}-\lambda_{1}^{2}}{18}},\quad P_{1}=\frac{\lambda_{1}}{3},\\
& \qquad \; P_{2}=\frac{-\lambda_{2}}{3};\\
&\textbf{CP4: }S=\frac{1}{2\sqrt{3}}\frac{2+\lambda_{2}^{2}
-\lambda_{1}^{2}}{\lambda_{1}^{2}-\lambda_{2}^{2}+1},\quad
U=\frac{1}{\sqrt{2}}\frac{\sqrt{\lambda_{1}^{2}
-\lambda_{2}^{2}+2}}{\lambda_{1}^{2}-\lambda_{2}^{2}+1}\\
&\qquad \; P_{1}=\frac{\lambda_{1}}{\lambda_{1}^{2}-\lambda_{2}^{2}+1},\quad
P_{2}=\frac{-\lambda_{2}}{\lambda_{1}^{2}-\lambda_{2}^{2}+1},\\
&\textbf{CP5: }S=\frac{1}{2\sqrt{3}}\frac{2+\lambda_{2}^{2}
-\lambda_{1}^{2}}{\lambda_{1}^{2}-\lambda_{2}^{2}+1},\quad
U=\frac{-1}{\sqrt{2}}\frac{\sqrt{\lambda_{1}^{2}
-\lambda_{2}^{2}+2}}{\lambda_{1}^{2}-\lambda_{2}^{2}+1}\\
&\qquad \; P_{1}=\frac{\lambda_{1}}{\lambda_{1}^{2}-\lambda_{2}^{2}+1},\quad
P_{2}=\frac{-\lambda_{2}}{\lambda_{1}^{2}-\lambda_{2}^{2}+1},\\
&\textbf{CP6: }U=0, \quad \left( \frac{S}{\sqrt{1/3}} \right)^{2}
+\left( \frac{P_{1}}{\sqrt{2/3}} \right)^{2}-\left( \frac{P_{2}}{\sqrt{2/3}} \right)^{2}=1.
\end{align*}
$\blacktriangledown$\; \textbf{CP1:}\\
CP1 lies in the KS region.
The eigenvalues of the Jacobian matrix for this CP are as $\varepsilon_{S}=\varepsilon_{P_{1}}=\varepsilon_{P_{2}}=-1/2$ and $\varepsilon_{U}=+1/2$. Thus, this point is a saddle point and is unstable both in the future and in the past.\\
This CP corresponds to the following unstable exact solution:
\begin{align*}
&\Theta=\frac{2}{t}, \quad \sigma=\frac{-1}{\sqrt{3}\;t},
\quad \psi_{1}=0, \quad \psi_{2}=0,\quad V=0\\
&\varphi_{1}=c_{1}, \quad \varphi_{2}=c_{2},
\end{align*}
where $c_{1}$ and $c_{2}$ are constants of integration.\\
The solution represents an expanding universe with a scale factor $a_{\mathrm{ave.}} \sim t^{2/3}$. Clearly, inflation cannot occur and since the scale factor expands at a slower rate than the horizon size, hence the horizon and flatness problems are not solved. The ratio $\sigma / \Theta \to \mathrm{constant}$ and so the model remains anisotropic.
$\blacktriangle$

$\blacktriangledown$ \; \textbf{CP2:}\\
According to the coordinate of this CP, it is in the physical region when $\lambda_{1}^{2} <6+\lambda_{2}^{2}$.
This CP belongs to the B-I universe and hence it lies on the hypersurface $W$ separating KS from B-III. Therefore, its neighborhood intersects all three types of universes.
In other words, starting from a point, near CP2, defined by
\begin{align*}
&S=\delta S, \quad U=\sqrt{\frac{6+\lambda_{2}^{2}-\lambda_{1}^{2}}{18}}
+\delta U, \quad P_{1}=\frac{\lambda_{1}}{3}+\delta P_{1},\\&
P_{2}=\frac{-\lambda_{2}}{3}+\delta P_{2},
\end{align*}
we can tell which universe this point belongs to by expanding the constraints equation (\ref{deq6}) to the first order:
\begin{align*}
\sqrt{\frac{2(6+\lambda_{2}^{2}-\lambda_{1}^{2})}{9}}+\frac{\lambda_{1}}{3}
\delta P_{1}-\frac{\lambda_{2}}{3}\delta P_{2}=-\frac{k}{b^{2}}.
\end{align*}
The solution can exist in either type of universe, depending on where the integration begins.\\

\begin{figure*}[ht]
\centering
\includegraphics[width=\textwidth]{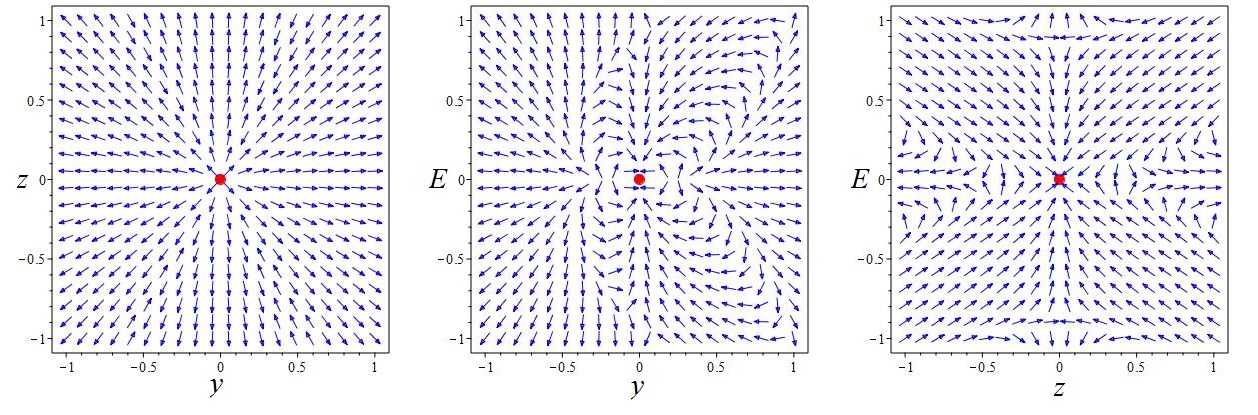}\\
\caption{This figure demonstrates three different Poincar\'{e} sections of the center manifold of CP2 in multiplicative mode. In plotting these sections of phase
portraits we have selected $\lambda_{2}=1$.}\label{3figs}
\end{figure*}

For this CP the linearization of the system has the following eigenvalues:
\begin{align*}
\varepsilon_{S}=\frac{\lambda_{1}^{2}-\lambda_{2}^{2}-2}{3},
\quad
\varepsilon_{U}=\varepsilon_{P_{1}}=\varepsilon_{P_{2}}=
\frac{\lambda_{1}^{2}-\lambda_{2}^{2}-6}{6}.
\end{align*}
So, when $2 < \lambda_{1}^{2}-\lambda_{2}^{2} <6$ the point is a physical saddle point ($\varepsilon_{S}>0$ and $\varepsilon_{U,P_{1},P_{2}}<0$). For the special cases $\lambda_{1}^{2}-\lambda_{2}^{2}=2$ and $\lambda_{1}^{2}-\lambda_{2}^{2}=6$, the behavior should be considered through center manifold theory:\\
1- Under the condition $\lambda_{1}^{2}-\lambda_{2}^{2}=2$, CP2 has
a one-dimensional (1d) center manifold. To consider its behavior, first of all, it is useful to study the system at CP2-origin. It means that we need to perform a transformation as $(S,U,P_{1},P_{2})|_{CP2} \rightarrow (s,u,p_{1},p_{2})=(0,0,0,0)$. In order to convert the system to a standard form, four new variables $(x,y,z,E)$ that are connected with $(s,u,p_{1},p_{2})$ via
\begin{align*}
s=& -\frac{\sqrt{3}}{2\lambda_{2}}x+E, \quad u=-\frac{1}{\sqrt{8}\lambda_{2}}x-\frac{\lambda_{2}}{\sqrt{8}}y
-\frac{\lambda_{1}}{\sqrt{8}}z,\\
p_{1}=&-\frac{\lambda_{1}}{\lambda_{2}}x+z,\quad p_{2}=x+y,
\end{align*}
are introduced. The center manifold is represented in the form
\begin{align*}
W^{c}=\{ (x,y,z,E)|\; |x|< \epsilon ,\; y=h_{1}(x),\;z=h_{2}(x),\;
E=h_{3}(x) \},
\end{align*}
where we assume that the mappings $h_{1}$, $h_{2}$, and $h_{3}$ are of the forms
\begin{align*}
h_{1}(x)&=c_{1}x^{2}+c_{2}x^{3}+\mathcal{O}(x^{5}),\\
h_{2}(x)&=c_{3}x^{2}+c_{4}x^{3}+\mathcal{O}(x^{5}),\\
h_{3}(x)&=c_{5}x^{2}+c_{6}x^{3}+\mathcal{O}(x^{5}),
\end{align*}
where $c_{i}$s are constant parameters. After some computations, we finally get the following topologically equivalent system:
\begin{align*}
x^{\prime}=&-\frac{3}{\lambda_{2}}x^{2}-\frac{9}{2\lambda_{2}^{2}}x^{3}
-\frac{5751}{64\lambda_{2}^{3}}x^{4}+\frac{891}{64\lambda_{2}^{4}}x^{5}
-\frac{3645}{8\lambda_{2}^{5}}x^{6}\\&+\frac{2187}{4\lambda_{2}^{6}}x^{7}
+\mathcal{O}(x^{8}),\\
y^{\prime}=&-\frac{2}{3}y,\\
z^{\prime}=&-\frac{2}{3}z,\\
E^{\prime}=&-\frac{2}{3}E.
\end{align*}
Therefore, the 1d center manifold is repeller and the nature of CP is again saddle.\\
2- For the special case $\lambda_{1}^{2}-\lambda_{2}^{2}=6$, the system has 3d center manifold. Transforming the system to the origin, i.e. $(S,U,P_{1},P_{2})|_{CP2} \rightarrow (s,u,p_{1},p_{2})=(0,0,0,0)$, and going to a new coordinate $(x,y,z,E)$ using
\begin{align*}
&s=-\frac{\sqrt{3}}{2\lambda_{2}}x+y,\quad u=E,\\
&p_{1}=-\frac{\sqrt{\lambda_{2}^{2}+6}}{\lambda_{2}}x
-\frac{\lambda_{2}}{\sqrt{\lambda_{2}^{2}+6}}z,\quad p_{2}=x+z,
\end{align*}
the system is cast in the standard form. Utilizing a 3d mapping as
\begin{align*}
h(y,z,E)=&c_{1}y^2+c_{2}z^2+c_{3}E^2+c_{4}yz+c_{5}yE+c_{6}zE\\&+c_{7}yzE
+c_{8}y^3+c_{9}z^3+c_{10}E^3+c_{11}yz^2\\&+c_{12}yE^2+c_{13}zE^2
+c_{14}zy^2\\&+c_{15}Ey^2+c_{16}Ez^2,
\end{align*}
where $c_{i}$s are constant parameters, one may reach a topologically equivalent system having the following form
\begin{align*}
x^{\prime}=&\frac{4}{3}x,\\
y^{\prime}=&-3yE^{2}+\cdots,\\
z^{\prime}=&-3zE^2+\cdots,\\
E^{\prime}=&\frac{-3}{2}E^{3}
+\frac{3}{2}Ey^{2}-\frac{9}{2\lambda_{2}^{2}+12}Ez^2+\cdots
\end{align*}
Therefore, even in this case, CP2 is again a saddle-node. Three different Poincar\'{e} sections of phase space of the center manifold have been demonstrated in fig.~\ref{3figs}.\\
CP2 transforms back to the following exact solution:
\begin{align*}
&\Theta=\frac{6}{\left( \lambda_{1}^{2}-\lambda_{2}^{2}  \right)t},
\quad \sigma =0,\\
&V=\frac{2\left( 6+\lambda_{2}^{2}-\lambda_{1}^{2} \right)}{\left(  \lambda_{1}^{2}-\lambda_{2}^{2}\right)^{2}t^{2}},\\
&\psi_{1}=\frac{2\lambda_{1}}{\left( \lambda_{1}^{2}-\lambda_{2}^{2}  \right)t},
\quad
\psi_{2}=\frac{-2\lambda_{2}}{\left( \lambda_{1}^{2}-\lambda_{2}^{2}  \right)t},\\
&\varphi_{1}=\frac{2\lambda_{1}}{\left( \lambda_{1}^{2}-\lambda_{2}^{2}  \right)}\ln (t)+c_{1},
\quad
\varphi_{2}=\frac{-2\lambda_{2}}{\left( \lambda_{1}^{2}-\lambda_{2}^{2}  \right)}\ln (t)+c_{2}.
\end{align*}
where $c_{1}$ and $c_{2}$ are constants of integration.\\
Isotropization occurs for every value of $\lambda_{1}$ and $\lambda_{2}$ in the sense that $\sigma / \Theta \to 0$ while existing inflation depends on the value of $\lambda_{1}$ and $\lambda_{2}$.\\
For CP2, inflation can happen in two ways:\\
1- Under the condition $\lambda_{1}^{2}-\lambda_{2}^{2} <2$ this solution violates the dominant energy condition and thus inflation occurs. The type of this inflation is power law as
\begin{align*}
a_{\mathrm{ave.}} \sim t^{\frac{2}{\lambda_{1}^{2}-\lambda_{2}^{2}}},
\end{align*}
But, in the aforementioned range, CP2 is an attractor point and hence it cannot be a good candidate because we cannot exit from it.\\
2- When $\lambda_{1}=\lambda_{2}$, we have exponential inflation as
\begin{align*}
a_{\mathrm{ave.}} \sim \exp \left( \frac{\Theta_{0} t}{3} \right),
\end{align*}
where $\Theta_{0}$ is an integration constant, but in this case, CP2 is again an attractor and the solution is asymptotically stable in the sense that solutions that start near it converge to it. Hence we put it aside.\\

If we restrict ourselves to the B-I universe, for the physical range $2 \leq \lambda_{1}^{2}-\lambda_{2}^{2} \leq 6$ which reflects saddle behavior, isotropy can be reached without inflation. Interestingly, this is consistent with Collins and Hawking's result that ordinary matter can achieve isotropy without inflation within the B-I universe~\cite{7}. Note that for $2 \leq \lambda_{1}^{2}-\lambda_{2}^{2} \leq 6$ the equation of state parameter is in the range $(-1/3)\leq \omega \leq 1$, so it contains the ordinary matter case ($0 \leq \omega <1$).
$\blacktriangle$

$\blacktriangledown$ \; \textbf{CP3:}\\
According to the value of $U$ and (\ref{d}), this CP corresponds to the contraction of the universe and therefore it contradicts the observational data.
$\blacktriangle$

$\blacktriangledown$ \; \textbf{CP4:}\\
According to the coordinate of this point, it is a physical point when we restrict ourselves to the condition $\lambda_{1}^{2}-\lambda_{2}^{2}>-1$. The eigenvalues of the linearization of our system in CP4 are:
\begin{align*}
\varepsilon_{1,2}&=\frac{\lambda_{2}^{2}-\lambda_{1}^{2}-2
\pm \sqrt{\left( 7\lambda_{2}^{2}-7\lambda_{1}^{2}+18 \right)\left(\lambda_{1}^{2}-\lambda_{2}^{2}+2  \right)}}{4\left(  \lambda_{1}^{2}-\lambda_{2}^{2}+1\right)},\\
\varepsilon_{3,4}&=\frac{\lambda_{2}^{2}-\lambda_{1}^{2}-2}{2\left(  \lambda_{1}^{2}-\lambda_{2}^{2}+1\right)}.
\end{align*}
Again the following exact solution after the pull-back of CP4 is obtained:
\begin{align*}
\Theta &=\frac{2\left(\lambda_{1}^{2}-\lambda_{2}^{2}+1\right)}
{\left(\lambda_{1}^{2}-\lambda_{2}^{2}\right)t}, \quad
V=\frac{2\left(\lambda_{1}^{2}-\lambda_{2}^{2}+2\right)}
{\left(\lambda_{1}^{2}-\lambda_{2}^{2}\right)^{2}t^{2}},\\
\sigma &=\frac{1}{\sqrt{3}}\frac{2+\lambda_{2}^{2}
-\lambda_{1}^{2}}{\left(\lambda_{1}^{2}-\lambda_{2}^{2}\right)t},\\
\psi_{1} &=\frac{2\lambda_{1}}
{\left(\lambda_{1}^{2}-\lambda_{2}^{2}\right)t}, \quad
\psi_{2} =\frac{-2\lambda_{2}}
{\left(\lambda_{1}^{2}-\lambda_{2}^{2}\right)t},\\
\varphi_{1} &=\frac{2\lambda_{1}}
{\left(\lambda_{1}^{2}-\lambda_{2}^{2}\right)}\ln (t)+c_{1}, \quad
\varphi_{2} =\frac{-2\lambda_{2}}
{\left(\lambda_{1}^{2}-\lambda_{2}^{2}\right)}\ln (t)+c_{2},
\end{align*}
where $c_{1}$ and $c_{2}$ are constants of integration.

Three types of universes are covered in the aforementioned domain:\\
1- Under the condition $-1<\lambda_{1}^{2}-\lambda_{2}^{2}<2$, this CP is in the B-III universe and all eigenvalues are real as $\varepsilon_{1}>0$ and $\varepsilon_{2,3,4}<0$ so the point is of saddle nature. Obviously, unlike inflation, isotropization cannot occur at this range because $\sigma / \Theta \to \mathrm{constant}$. If one sets $\lambda_{1}=\lambda_{2}$, then exponential inflation as
\begin{align*}
a_{\mathrm{ave.}} \sim \exp \left( \frac{\Theta_{0} t}{3} \right),
\end{align*}
is reached, while for $0< \lambda_{1}^{2}-\lambda_{2}^{2}<2$, the power-law inflation happens:
\begin{align*}
a_{\mathrm{ave.}} \sim t^{\frac{2\left( \lambda_{1}^{2}-\lambda_{2}^{2}+1 \right)}{3\left( \lambda_{1}^{2}-\lambda_{2}^{2}\right)}}.
\end{align*}
The equation of state parameter takes the values $\omega_{\mathrm{eff.}}=-1$ and $-1<\omega_{\mathrm{eff.}} <-1/3$, respectively. Both refer to an accelerated type of expansion.\\

2- For $\lambda_{1}^{2}-\lambda_{2}^{2}=2$ the CP lies on the B-I universe. Isotropy can be reached without inflation. The eigenvalues of the Jacobian matrix for this CP becomes $\varepsilon_{1}=0$, $\varepsilon_{2}=\varepsilon_{3}=\varepsilon_{4}=-2/3$. So we must consider it through center manifold theory. Again, let us consider the problem at CP-origin meaning that the system is transformed into a coordinate $(s,u,p_{1},p_{2})$ so that the position of the CP becomes $(0,0,0,0)$. Then using
\begin{align*}
s&=\frac{-\sqrt{3}}{2\lambda_{2}}x+E, \quad p_{1}=\frac{\sqrt{\lambda_{2}^{2}+2}}{-\lambda_{2}}x+z,\\
u&=\frac{-\sqrt{2}}{4\lambda_{2}}x-\frac{\sqrt{2}\lambda_{2}}{4}y
-\frac{\sqrt{2\lambda_{2}^{2}+4}}{4}z,\\
p_{2}&=x+y,
\end{align*}
the system is presented in the new coordinate $(x,y,z,E)$. After utilizing the mappings
\begin{align*}
h_{1}(x)&=c_{1}x^{2}+c_{2}x^{3}+\mathcal{O}(x^{5}),\\
h_{2}(x)&=c_{3}x^{2}+c_{4}x^{3}+\mathcal{O}(x^{5}),\\
h_{3}(x)&=c_{5}x^{2}+c_{6}x^{3}+\mathcal{O}(x^{5}),
\end{align*}
we get the following system which is topologically equivalent to the system at the CP4-origin:
\begin{align*}
x^{\prime}=&-\frac{3}{\lambda_{2}}x^{2}-\frac{9}{2\lambda_{2}^{2}}x^{3}
-\frac{5751}{64\lambda_{2}^{3}}x^{4}+\frac{891}{64\lambda_{2}^{4}}x^{5}
-\frac{3645}{8\lambda_{2}^{5}}x^{6}\\&+\frac{2187}{4\lambda_{2}^{6}}x^{7}
+\mathcal{O}(x^{8}),\\
y^{\prime}=&-\frac{2}{3}y,\\
z^{\prime}=&-\frac{2}{3}z,\\
E^{\prime}=&-\frac{2}{3}E.
\end{align*}
Therefore, the one-dimensional center manifold is repeller and the nature of CP under the aforementioned condition is saddle. Thus it is acceptable and again is consistent with Collins and Hawking's result. \\

3- In the range $2<\lambda_{1}^{2}-\lambda_{2}^{2}< 18/7$ all eigenvalues are real and negative, while for $\lambda_{1}^{2}-\lambda_{2}^{2}> 18/7$, $\varepsilon_{1}$ and $\varepsilon_{2}$ become complex with the same negative real part. So CP4 is asymptotically stable (attractor) for $\lambda_{1}^{2}-\lambda_{2}^{2}>2$. In this range, the CP belongs to the KS universe and neither inflation nor isotropization occurs at it.
$\blacktriangle$

$\blacktriangledown$ \; \textbf{CP5:}\\
In this CP, the physical range in which we have an expanding universe is $-2 \leq \lambda_{1}^{2}-\lambda_{2}^{2} <-1$. All discussions performed in this range for CP4 are valid for this CP because of the same position values.
$\blacktriangle$

$\blacktriangledown$ \; \textbf{CP6:}\\
CP6 identifies a family of critical points so that the coordinates $(S,P_{1},P_{2})$ form a hyperboloid of one sheet or so-called hyperbolic hyperboloid. Indeed, they form a connected surface, which has a negative Gaussian curvature at every point. This implies near every point the intersection of the hyperboloid and its tangent plane at the point consists of two branches of curve that have distinct tangents at the point. In our case, these branches of curves are lines and thus this surface is a doubly ruled surface.\\
The pull-back of points out of CP6 gives an exact solution of the form
\begin{align*}
&\Theta = \frac{1}{t}, \quad \sigma = \frac{S_{0}}{t}, \quad \psi_{1}=\frac{P_{10}}{t},
\quad \psi_{2}=\frac{P_{20}}{t}, \quad V=0,\\
&\varphi_{1}(t)=P_{10} \ln (t)+c_{1},
\quad \varphi_{2}(t)=P_{20} \ln (t)+c_{2},
\end{align*}
where $c_{1}$ and $c_{2}$ are constants of integration.\\
All the points of this family belong to the B-I universe and never inflation can occur at it because $a_{\mathrm{ave.}} \sim t^{1/3}$ and $\omega_{\mathrm{eff.}}=1$. The scale factor expands at a slower rate than the horizon size, hence the horizon and flatness problems are not solved.\\

For the linearization of system around a given point $(S_{0},0,P_{10},P_{20})$ we find the following eigenvalues:
\begin{align*}
&\varepsilon_{1}=1-\frac{\lambda_{1}P_{10}}{2}-\frac{\lambda_{2}P_{20}}{2},
\\
&\varepsilon_{2}=\frac{\sqrt{18\left( 2-3P_{10}^{2}+3P_{20}^{2} \right)}}{9}
+\frac{4}{3}=\frac{2}{\sqrt{3}}S_{0}+\frac{4}{3},
\\
&\varepsilon_{3}=\varepsilon_{4}=0.
\end{align*}
Investigating this two-dimensional center manifold is somewhat difficult hence let us investigate two sample points as $(P_{10},P_{20})=(1,1)$ and $(P_{10},P_{20})=(1,1/\sqrt{3})$ to show what the whole is like. Unlike the former one, for the latter one we have isotropization.

$\maltese$ \textbf{The case $(P_{10},P_{20})=(1,1)$ (without isotropization):}\\
If the system is transformed as $(S,U,P_{1},P_{2})|_{\mathrm{CP6}} \rightarrow (s,u,p_{1},p_{2})=(0,0,0,0)$ and then we use the following transformations,
\begin{align*}
&s=\frac{\sqrt{3}}{2}\left( y+z-E \right),\\
&u=x,\quad p_{1}=y+E, \quad p_{2}=y+z,
\end{align*}
we arrive at the following topologically equivalent system
\begin{align*}
x^{\prime}=&\left( 1-\frac{\lambda_{1}}{2}-\frac{\lambda_{2}}{2} \right)x,\\
y^{\prime}=&2y,\\
z^{\prime}=&\frac{-1}{4}z^{4}+\frac{5}{4}zE^{3}+\frac{7}{4}Ez^{3}
-\frac{11}{4}z^{2}E^{2}+\frac{3}{32}z^{5}-\frac{9}{8}Ez^{4}\\&
+\frac{69}{16}z^{3}E^{2}-\frac{45}{8}E^{3}z^{2}+\frac{75}{32}zE^{4},\\
E^{\prime}=&\frac{5}{4}E^{4}-\frac{1}{4}Ez^{3}-\frac{11}{4}zE^{3}
+\frac{7}{4}z^{2}E^{2}+\frac{75}{32}E^{5}-\frac{45}{8}zE^{4}\\
&+\frac{69}{16}z^{2}E^{3}-\frac{9}{8}E^{2}z^{3}+\frac{3}{32}Ez^{4},
\end{align*}
where we have exploited the following mappings:
\begin{align*}
h_{1}(z,E)=&c_{1}z^{2}+c_{2}zE+c_{3}E^{2},\\
h_{2}(z,E)=&c_{4}z^{2}+c_{5}zE+c_{6}E^{2}.
\end{align*}
\begin{figure}[ht]
\centering
\includegraphics[width=10cm,height=10cm]{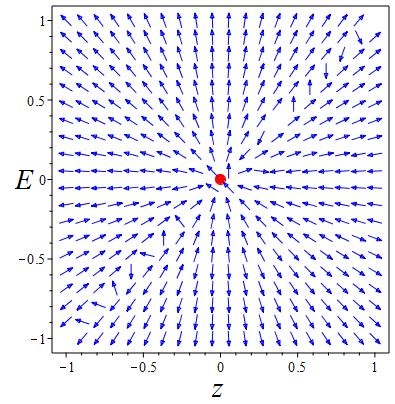}\\
\caption{This figure demonstrates phase portrait of 2d center manifold of CP6 of multiplicative mode for the case $(P_{10},P_{20})=(1,1)$.}\label{fig1}
\end{figure}
\begin{figure}[ht]
\centering
\includegraphics[width=10cm,height=10cm]{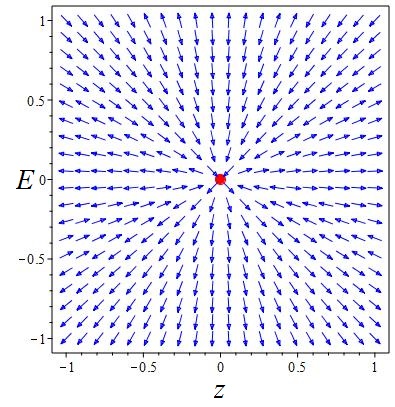}\\
\caption{This figure indicates phase portrait of 2d center manifold of CP6 of multiplicative mode for the case $(P_{10},P_{20})=(1,1/\sqrt{3})$.}\label{fig2}
\end{figure}
Thus the system at CP6 has a saddle nature which is physical. The behavior of 2d center manifold has been shown in fig.~\ref{fig1}. The saddle nature of the center manifold is obvious from this portrait.

$\maltese$ \textbf{The case $(P_{10},P_{20})=(1,1/\sqrt{3})$ (with isotropization):}\\
The interesting property of this case is that the model isotropizes.\\
Performing a transformation as $(S,U,P_{1},P_{2})|_{\mathrm{CP6}} \rightarrow (s,u,p_{1},p_{2})=(0,0,0,0)$ and then applying the transformation
\begin{align*}
&s=\frac{1}{2}y+E, \quad u=x,\\
&p_{1}=\sqrt{3}\; y+\frac{1}{\sqrt{3}}z, \quad p_{2}=y+z,
\end{align*}
our system will be expressed in a new coordinate $(x,y,z,E)$. By the use of two mappings
\begin{align*}
h_{1}(z,E)=&c_{1}z^{2}+c_{2}zE+c_{3}E^{2},\\
h_{2}(z,E)=&c_{4}z^{2}+c_{5}zE+c_{6}E^{2},
\end{align*}
we arrive at a system that is topologically equivalent to the original system:
\begin{align*}
&x^{\prime}=\left( 1-\frac{\lambda_{1}}{2}-\frac{\lambda_{2}}{2\sqrt{3}} \right)x,\\
&y^{\prime}=\frac{4}{3}y,\\
&z^{\prime}=-\sqrt{3}\;zE^3+\frac{1}{\sqrt{3}}Ez^3+\frac{15}{8}zE^4
-\frac{5}{4}z^{3}E^{2}+\frac{5}{24}z^{5},\\
&E^{\prime}=-\sqrt{3}\;E^{4}+\frac{1}{\sqrt{3}}z^{2}E^{2}+\frac{5}{24}Ez^{4}
-\frac{5}{4}z^{2}E^{3}+\frac{15}{8}E^{5}.
\end{align*}
As is clear from fig.~\ref{fig2}, this 2d center manifold is also saddle at the origin and thus the point is physical.
$\blacktriangle$

\subsection{Class-2: Collective mode: $V(\varphi_{1},\varphi_{2})=V_{1}(\varphi_{1})+V_{2}(\varphi_{2})$}

In this subsection, we study the potential of the form $V(\varphi_{1},\varphi_{2})=V_{1}(\varphi_{1})+V_{2}(\varphi_{2})=V_{01}\exp (-\lambda_{1}\varphi_{1})+V_{02}\exp (-\lambda_{2}\varphi_{2})$.\\
Defining the variables $S$, $U_{1}$, $U_{2}$, $P_{1}$ and $P_{2}$ by
\begin{align}
\label{d2}
S=\frac{\sigma}{\Theta}, \quad U_{1}=\frac{\sqrt{V_{1}}}{\Theta}, \quad
U_{2}=\frac{\sqrt{V_{2}}}{\Theta}, \quad
P_{1}=\frac{\psi_{1}}{\Theta}, \quad P_{2}=\frac{\psi_{2}}{\Theta},
\end{align}
the eqs.~(\ref{pdeq1})-(\ref{pdeq6}) turn out to be
\begin{align}
\label{1deq1}
\Theta^{\prime}=&\Theta \biggl( -\frac{1}{3}-2S^{2}+U_{1}^{2}+U_{2}^{2}
-P_{1}^{2}+P_{2}^{2} \biggl),\\
\label{1deq2}
S^{\prime}=&\frac{-1}{3\sqrt{3}}-\frac{2}{3}S+\frac{1}{\sqrt{3}}S^{2}
+2S^{3}+P_{1}^{2}\biggl( \frac{1}{2\sqrt{3}}+S \biggl)\nonumber\\&
-P_{2}^{2}\biggl( \frac{1}{2\sqrt{3}}+S \biggl)
+\left(U_{1}^{2}+U_{2}^{2} \right)\biggl( \frac{1}{\sqrt{3}}-S \biggl),\\
\label{1deq3}
U_{1}^{\prime}=&U_{1}\biggl( \frac{1}{3}+2S^{2}-U_{1}^{2}-U_{2}^2-\frac{\lambda_{1}}{2}
P_{1}+P_{1}^{2}-P_{2}^{2} \biggl),\\
\label{1deq4}
U_{2}^{\prime}=&U_{2}\biggl( \frac{1}{3}+2S^{2}-U_{1}^{2}-U_{2}^2-\frac{\lambda_{2}}{2}
P_{2}+P_{1}^{2}-P_{2}^{2} \biggl),\\
\label{1deq5}
P_{1}^{\prime}=&P_{1}\biggl( -\frac{2}{3}+2S^{2}+P_{1}^{2}-P_{2}^{2} \biggl)
-P_{1}\left( U_{1}^{2}+U_{2}^{2} \right)+\lambda_{1}U_{1}^{2},\\
\label{1deq6}
P_{2}^{\prime}=&P_{2}\biggl( -\frac{2}{3}+2S^{2}+P_{1}^{2}-P_{2}^{2} \biggl)
-P_{2}\left( U_{1}^{2}+U_{2}^{2} \right)-\lambda_{2}U_{2}^{2},
\end{align}
and the constraint equation becomes
\begin{align}
\label{1deq7}
S^{2}+U_{1}^{2}+U_{2}^{2}+\frac{1}{2}P_{1}^{2}-\frac{1}{2}P_{2}^{2}+\frac{k}{b^{2}}
=\frac{1}{3}.
\end{align}
For the above system, seventeen critical points and one family of critical points are obtained, but since we have concentrated on the expanding universe, eight CPs are removed and the remains are:
\begin{align*}
&\textbf{CP1: }S=\frac{-1}{2\sqrt{3}}, \quad U_{1}=0, \quad U_{2}=0,
\quad P_{1}=0, \quad P_{2}=0,\\
&\textbf{CP2: }S=0, \quad U_{1}=0,
\quad U_{2}=\sqrt{\frac{\lambda_{2}^{2}+6}{18}},
\quad P_{1}=0,\\&\qquad \quad P_{2}=\frac{-\lambda_{2}}{3},\\
&\textbf{CP3: }S=0, \quad U_{1}=\sqrt{\frac{6-\lambda_{1}^{2}}{18}},
\quad U_{2}=0, \quad P_{1}=\frac{\lambda_{1}}{3}, \\
& \qquad \quad P_{2}=0,\\
&\textbf{CP4: }S=\frac{1}{2\sqrt{3}}\frac{2-\lambda_{1}^{2}}{\lambda_{1}^{2}+1},
\quad U_{1}=\frac{1}{\sqrt{2}}\frac{\sqrt{\lambda_{1}^{2}+2}}{\lambda_{1}^{2}+1},
\quad U_{2}=0,\\ & \qquad \quad P_{1}=\frac{\lambda_{1}}{\lambda_{1}^{2}+1},
\quad P_{2}=0,\\
&\textbf{CP5: }S=\frac{1}{2\sqrt{3}}\frac{2+\lambda_{2}^{2}}{1-\lambda_{2}^{2}},
\quad U_{1}=0, \quad U_{2}=\frac{1}{\sqrt{2}}\frac{\sqrt{2-\lambda_{2}^{2}}}{\lambda_{2}^{2}-1},\\ & \qquad \quad P_{1}=0, \quad P_{2}=\frac{\lambda_{2}}{\lambda_{2}^{2}+1},\\
&\textbf{CP6: }S=\frac{1}{2\sqrt{3}}\frac{2+\lambda_{2}^{2}}{1-\lambda_{2}^{2}},
\quad U_{1}=0, \quad U_{2}=\frac{-1}{\sqrt{2}}\frac{\sqrt{2-\lambda_{2}^{2}}}{\lambda_{2}^{2}-1},\\ & \qquad \quad P_{1}=0, \quad P_{2}=\frac{\lambda_{2}}{\lambda_{2}^{2}+1},\\
&\textbf{CP7: }S=0, \quad U_{1}=-\, \frac{\lambda_{2}}{3\sqrt{2}}
\frac{\sqrt{6\lambda_{2}^{2}-6\lambda_{1}^{2}
-\lambda_{1}^{2}\lambda_{2}^{2}}}{\lambda_{1}^{2}-\lambda_{2}^{2}},\\
&\qquad \quad U_{2}= \frac{\lambda_{1}}{3\sqrt{2}}
\frac{\sqrt{\lambda_{1}^{2}\lambda_{2}^{2}+6\lambda_{1}^{2}
-6\lambda_{2}^{2}}}{\lambda_{1}^{2}-\lambda_{2}^{2}},\\
&\qquad \quad P_{1}=\frac{\lambda_{1}\lambda_{2}^{2}}{3\left( \lambda_{2}^{2}-\lambda_{1}^{2} \right)},\quad P_{2}=\frac{\lambda_{1}^{2}\lambda_{2}}{3\left( \lambda_{2}^{2}-\lambda_{1}^{2} \right)},\\
&\textbf{CP8: } S=\frac{-1}{2\sqrt{3}}\frac{\lambda_{1}^{2}
\lambda_{2}^{2}+2\lambda_{1}^{2}-2\lambda_{2}^{2}}{\lambda_{1}^{2}\lambda_{2}^{2}
-\lambda_{1}^{2}+\lambda_{2}^{2}},\\
&U_{1}= \frac{\lambda_{2}}{\sqrt{2}}
\frac{\sqrt{\lambda_{1}^{2}\lambda_{2}^{2}
-2\lambda_{1}^{2}+2\lambda_{2}^{2}}}{
\lambda_{1}^{2}\lambda_{2}^{2}
-\lambda_{1}^{2}+\lambda_{2}^{2}}\\
&U_{2}=-\, \frac{\lambda_{1}}{\sqrt{2}} \frac{\sqrt{2\lambda_{1}^{2}-2\lambda_{2}^{2}
-\lambda_{1}^{2}\lambda_{2}^{2}}}{\lambda_{1}^{2}\lambda_{2}^{2}
-\lambda_{1}^{2}+\lambda_{2}^{2}},\\
&P_{1}=\frac{\lambda_{1}\lambda_{2}^{2}}{\lambda_{1}^{2}\lambda_{2}^{2}
-\lambda_{1}^{2}+\lambda_{2}^{2}},
\quad
P_{2}=\frac{\lambda_{1}^{2}\lambda_{2}}{\lambda_{1}^{2}\lambda_{2}^{2}
-\lambda_{1}^{2}+\lambda_{2}^{2}},\\
&\textbf{CP9: }U_{1}=0, \quad U_{2}=0,
\\& \left( \frac{S}{\sqrt{1/3}} \right)^{2}
+\left( \frac{P_{1}}{\sqrt{2/3}} \right)^{2}-\left( \frac{P_{2}}{\sqrt{2/3}} \right)^{2}=1.
\end{align*}
$\blacktriangledown$\; \textbf{CP1:}\\
CP1 belongs to KS universe. The eigenvalues of it show that it is a saddle node: $\varepsilon_{S}=\varepsilon_{P_{1}}=\varepsilon_{P_{2}}=-1/2$ and $\varepsilon_{U_{1}}=\varepsilon_{U_{2}}=+1/2$.\\
This CP corresponds to the following unstable exact solution:
\begin{align*}
&\Theta =\frac{2}{t}, \quad \sigma = \frac{-1}{\sqrt{3}\; t},
\quad \psi_{1}=\psi_{2}=0,\\
&V_{1}=V_{2}=0, \quad \varphi_{1}=c_{1},\quad \varphi_{2}=c_{2},
\end{align*}
where $c_{1}$ and $c_{2}$ are constants of integration.\\
According to the exact solution, neither inflation nor isotropization occurs at this CP. The solution at this point represents an expanding universe with a scale factor of the form $a_{\mathrm{ave.}} \sim t^{2/3}$. The scale factor expands at a slower rate than the horizon size and so the horizon and flatness problems are not solved.
$\blacktriangle$

$\blacktriangledown$\; \textbf{CP2:}\\
CP2 is located in the B-I region.\\
CP2 transforms back to the following exact solution:
\begin{align*}
&\Theta=\frac{-6}{\lambda_{2}^{2}t}, \quad \sigma =0, \quad \psi_{1}=0,
\quad \psi_{2}=\frac{2}{\lambda_{2}t},\\
&V_{1}=0, \quad V_{2}=\frac{2\left( \lambda_{2}^{2}+6 \right)}{\lambda_{2}^{4}t^{2}}, \quad \varphi_{1}=c_{1}, \quad\varphi_{2}=\frac{2\ln (t)}{\lambda_{2}}+c_{2},
\end{align*}
where $c_{1}$ and $c_{2}$ are constants of integration.\\
Unlike isotropization, inflation cannot occur at this CP because the scale factor evolves as $a_{\mathrm{ave.}} \sim t^{-2/\lambda_{2}^{2}}$. Clearly, if $\lambda_{2}$ be a purely imaginary number, then under the condition $|\lambda_{2}|<\sqrt{2}$ inflation can successfully be obtained. In this case, $\varphi_{2}$ becomes purely imaginary, and hence the potential remains real. Therefore, if we want to have inflation in this CP, one of the scalar fields, $\varphi_{2}$, must be a purely imaginary function. At this point, the amount of EoS parameter is $-1-\lambda_{2}^{2}/3$ indicating an accelerated expansion era for a suitable value of $\lambda_{2}$.\\
The eigenvalues of the linearized system indicate saddle nature for purely imaginary values of $\lambda_{2}$ when $|\lambda_{2}|<\sqrt{2}$ which is physically admissible:
\begin{align*}
&\varepsilon_{P_{1}}=\frac{-\lambda_{2}^{2}}{6},\quad \varepsilon_{P_{2}}=\frac{-\left( \lambda_{2}^{2}+2 \right)}{3}, \\
&\varepsilon_{S}=\varepsilon_{U_{1}}=\varepsilon_{U_{2}}=\frac{-\left( \lambda_{2}^{2}+6 \right)}{6}.
\end{align*}
As is clear, for real values of $\lambda_{2}$, CP2 is an attractor so it is not suitable since we cannot exit from it. Even, if $V_{2}$ treats as a cosmological constant, i.e. $\lambda_{2}=0$, then by studying the center manifold it becomes clear that it has again attractor nature. In a nutshell, the result of investigating through the center manifold theory is as follows:
\begin{align*}
x^{\prime}=&-x,\\
y^{\prime}=&-y,\\
z^{\prime}=&-z,\\
E^{\prime}=&-\,\frac{\lambda_{1}^{2}}{2}E^{3}+\frac{4\lambda_{1}^{2}-3}{4}E^{5},\\
R^{\prime}=&-\,\frac{2}{3}R,
\end{align*}
where we have moved the system to the origin, $(S,U_{1},U_{2},P_{1},P_{2})|_{\mathrm{CP2}} \to (s,u_{1},u_{2},p_{1},p_{2})=(0,0,0,0,0)$, and then we have applied the following transformations:
\begin{align*}
s=z+2R, \quad u_{1}=E, \quad u_{2}=R, \quad p_{1}=y, \quad p_{2}=x.
\end{align*}
It should be mentioned that for $\lambda_{2}=0$ we have exponential inflation as $a_{\mathrm{ave.}} \sim \exp (\Theta_{0}t/3)$ with $\omega_{\mathrm{eff.}}=-1$.
$\blacktriangle$

$\blacktriangledown$\; \textbf{CP3:}\\
This case is very similar to CP2 except that we must apply the following changes in the scale factor, eigenvalues, EoS parameter, and exact solutions:
\begin{align*}
&\lambda_{2} \rightarrow \lambda_{1},\\
&\lambda_{2}^{2} \rightarrow -\lambda_{1}^{2}.
\end{align*}
The outcome is that we have a CP in the B-I universe which for $0<\lambda_{1}<\sqrt{2}$ the solution violates the dominant energy condition and thus inflation happens. Since the scale factor grows faster than the horizon this model solves the horizon and flatness problems.
The ratio $\sigma /\Theta \to 0$ as the equilibrium point is approached and the model can be said to approach isotropy as $t \to \infty$. Under condition $0<\lambda_{1}<\sqrt{2}$, the point is a saddle-node that lies on the accelerated era and thus is physically suitable.\\
For $\lambda_{1}=0$ we have exponential inflation as $a_{\mathrm{ave.}} \sim \exp (\Theta_{0}t/3)$ with $\omega_{\mathrm{eff.}}=-1$.\\
As a final point, it should be mentioned that this CP is in the physical region if $\lambda_{1} \leq \sqrt{6}$ otherwise $V_{1}$ must be a purely imaginary function of the scalar field $\varphi_{1}$ which is strange.
$\blacktriangle$

$\blacktriangledown$\; \textbf{CP4:}\\
The eigenvalues of the linearization of the system in CP4 are:
\begin{align*}
&\varepsilon_{1}=\frac{\lambda_{1}^{2}}{2\left( \lambda_{1}^{2}+1 \right)}, \quad \varepsilon_{2}=\varepsilon_{3}=-\,\frac{\lambda_{1}^{2}+2}{2\left( \lambda_{1}^{2}+1 \right)},\\
&\varepsilon_{4,5}=\frac{-\left( \lambda_{1}^{2}+2 \right)\pm \sqrt{-7\lambda_{1}^{4}+4\lambda_{1}^{2}+36}}{4\left( \lambda_{1}^{2}+1 \right)}.
\end{align*}
Again, we find the following exact solution after the pull-back of CP4:
\begin{align*}
&\Theta =\frac{2\left( \lambda_{1}^{2}+1 \right)}{\lambda_{1}^{2}t}, \quad \sigma =\frac{\sqrt{3}\left( \lambda_{1}^{2}-2 \right)}{3\lambda_{1}^{2}t}, \quad \psi_{1}=\frac{2}{\lambda_{1}t},\\
&\psi_{2}=0, \quad V_{1}=\frac{2\left( \lambda_{1}^{2}+2 \right)}{\lambda_{1}^{4}t^{2}},\quad V_{2}=0,\\
&\varphi_{1}=\frac{2\ln (t)}{\lambda_{1}}+c_{1}, \quad \varphi_{2}=c_{2},
\end{align*}
where $c_{1}$ and $c_{2}$ are constants of integration.\\
Three types of universes are covered in this CP as follows:\\
1- If $0\leq \lambda_{1} <\sqrt{2}$, the CP falls in the B-III universe with the EoS $-1 \leq \omega_{\mathrm{eff.}}<-1/3$ which denotes an accelerated era. For $0< \lambda_{1} <\sqrt{2}$ and $\lambda_{1}=0$, inflation occurs with power-law and exponential laws as $a_{\mathrm{ave.}} \sim t^{2(\lambda_{1}^{2}+1)/(3\lambda_{1}^{2})}$ and $a_{\mathrm{ave.}} \sim \exp (\Theta_{0}t/3)$, respectively. Consequently, the horizon and flatness problems are solved.
The universe does not isotropize since the ratio of shear to expansion factor remains constant.
For $\lambda_{1}< \sqrt{2}$ all eigenvalues are real and represent a saddle-node which is desirable.\\
2- For $\lambda_{1} = \sqrt{2}$, CP4 is a saddle-node that belongs to the B-I universe with the properties that inflation does not happen while isotropization occurs in the sense that $\sigma / \Theta \to 0$. In this case, we have $\omega_{\mathrm{eff.}}=-1/3$.\\
3- For $\lambda_{1} > \sqrt{2}$, CP4 lies on KS universe with $\omega_{\mathrm{eff.}} >-1/3$. Neither inflation nor isotropization happens at this CP. In the range $\sqrt{2}<\lambda_{1} \leq \sqrt{18/7}$, all eigenvalues are real and the point is a saddle-node while for $\lambda_{1} > \sqrt{18/7}$, $\varepsilon_{4}$ and $\varepsilon_{5}$ become complex with the same negative real part and the point is a saddle spiral.
$\blacktriangle$

$\blacktriangledown$\; \textbf{CP5:}\\
This point is in physical region when $\lambda_{2} \in [0,1) \cup (1,\sqrt{2}]$. In the range $\lambda_{2} \in [0,1) \cup (1,\sqrt{2})$ CP5 belongs to the B-III universe while for $\lambda_{2}=\sqrt{2}$, it falls in the B-I universe. According to its eigenvalues,
\begin{align*}
&\varepsilon_{1}=\frac{\lambda_{2}^{2}}{2\left( \lambda_{2}^{2}-1 \right)}, \quad \varepsilon_{2}=\varepsilon_{3}=\frac{-\lambda_{2}^{2}+2}{2\left( \lambda_{2}^{2}-1 \right)},\\
&\varepsilon_{4,5}=\frac{- \lambda_{2}^{2}+2 \pm \sqrt{-7\lambda_{2}^{4}-4\lambda_{2}^{2}+36}}{4\left( \lambda_{2}^{2}-1 \right)},
\end{align*}
this point in the aforementioned region has a saddle nature.\\
This CP corresponds to the following unstable exact solution:
\begin{align*}
&\Theta =\frac{2\left( \lambda_{2}^{2}-1 \right)}{\lambda_{2}^{2}t},
\quad \sigma =\frac{-1}{\sqrt{3}}\frac{\lambda_{2}^{2}+2}{\lambda_{2}^{2}t},
\quad \psi_{1}=0,\\
&\psi_{2}=\frac{2}{\lambda_{2}t}, \quad V_{1}=0, \quad V_{2}=\frac{2\left(  2-\lambda_{2}^{2}\right)}{\lambda_{2}^{4}t^{2}},\\
&\varphi_{1}=c_{1}, \quad \varphi_{2}=\frac{2\ln (t)}{\lambda_{2}}+c_{2}.
\end{align*}
where $c_{1}$ and $c_{2}$ are constants of integration.\\
In both cases (i.e. B-I and B-III), neither the corresponding universe experiences inflation nor isotropizes for a positive value of $\lambda_{2}$. However for $\lambda_{2}=0$ we have exponential inflation with $\omega_{\mathrm{eff.}}=-1$.\\
If we allow $\lambda_{2}$ to be a purely imaginary number then $\varphi_{2}$ must be a purely imaginary function and consequently, potential $V_{2}$ remains a positive real function. In this case, for $0<|\lambda_{2}|<\sqrt{2}$ and $|\lambda_{2}|=\sqrt{2}$ power-law inflation, $a_{\mathrm{ave.}} \sim t^{2(\lambda_{2}^{2}-1)/(3\lambda_{2}^{2})}$, and isotropization happens at this CP. The former one occurs in the B-III background with $-1 <\omega_{\mathrm{eff.}}<-1/3$ while the latter one happens in the B-I with $\omega_{\mathrm{eff.}}=1$. In the B-III universe, since the scale factor grows faster than the horizon hence this model solves the horizon and flatness problems.
$\blacktriangle$

$\blacktriangledown$\; \textbf{CP6:}\\
This CP is very similar to CP5 and all discussions are valid here except that the physical region compresses to $0 \leq \lambda_{2} <1$ for maintaining the expansion of the universe because of the negative sign in $U_{2}$. The eigenvalues and exact solutions are also the same.
$\blacktriangle$

$\blacktriangledown$\; \textbf{CP7:}\\
CP7 is a point in the B-I universe.
According to radical terms in $U_{1}$ and $U_{2}$, to have positive convex potentials, one of the $\lambda_{1}$ or $\lambda_{2}$ must be a purely imaginary number; for example, let $\lambda_{2}=ci$ where $c\in \mathbb{R}$ and $i=\sqrt{-1}$. This makes one of the scalar fields, $\varphi_{2}$, a purely imaginary function. In this case, pawer-law inflation, $a_{\mathrm{ave.}} \sim t^{2(\lambda_{2}^{2}-\lambda_{1}^{2})/(\lambda_{1}^{2}\lambda_{2}^{2})}$, can be achieved by restricting ourselves to $0<c<\sqrt{2}\lambda_{1}/\sqrt{\lambda_{1}^2-2}$ and $\lambda_{1}>\sqrt{2}$.  Since the scale factor grows faster than the horizon this model solves the horizon and flatness problems.
Exponential inflation is also attainable when $\lambda_{1}=0$ or $c=0$.\\
The pull-back of points out of CP7 gives an exact solution of the form:
\begin{align*}
&\Theta =\frac{6\left( \lambda_{2}^{2}-\lambda_{1}^{2} \right)}{\lambda_{1}^{2}\lambda_{2}^{2}t}, \quad \sigma =0, \quad \psi_{1}=\frac{2}{\lambda_{1}t},\\
&\psi_{2}=\frac{2}{\lambda_{2}t}, \quad V_{1}=\frac{12\lambda_{2}^{2}-2\lambda_{1}^{2}\left( \lambda_{2}^{2}+6 \right)}{\lambda_{1}^{4}\lambda_{2}^{2}t^{2}},\\
&V_{2}=\frac{-12\lambda_{2}^{2}+2\lambda_{1}^{2}\left( \lambda_{2}^{2}+6 \right)}{\lambda_{1}^{2}\lambda_{2}^{4}t^{2}},\\
&\varphi_{1}=\frac{2\ln (t)}{\lambda_{1}}+c_{1}, \quad \varphi_{2}=\frac{2\ln (t)}{\lambda_{2}}+c_{2},
\end{align*}
where $c_{1}$ and $c_{2}$ are constants of integration.
The model has $\sigma / \Theta \to 0$ and so will isotropize.\\
The eigenvalues of the linearization of the system at CP7 are
\begin{align*}
&\varepsilon_{1}=-\,\frac{\lambda_{1}^{2}\lambda_{2}^{2}
+2\lambda_{1}^{2}-2\lambda_{2}^{2}}{3\left( \lambda_{1}^{2}-\lambda_{2}^{2} \right)},\\ &\varepsilon_{2}=\varepsilon_{3}=-\,\frac{\lambda_{1}^{2}
\lambda_{2}^{2}+6\lambda_{1}^{2}-6\lambda_{2}^{2}}{6\left( \lambda_{1}^{2}-\lambda_{2}^{2} \right)},\\
&\varepsilon_{4,5}=\frac{-\lambda_{1}^{2}\lambda_{2}^{2}
-6\lambda_{1}^{2}+6\lambda_{2}^{2}\pm \sqrt{A}}{12\left( \lambda_{1}^{2}-\lambda_{2}^{2} \right)},
\end{align*}
where
\begin{align*}
A=9\lambda_{1}^{4}\lambda_{2}^{4}+60\lambda_{1}^{4}\lambda_{2}^{2}
-60\lambda_{1}^{2}\lambda_{2}^{4}+36\lambda_{1}^{4}
-72\lambda_{1}^{2}\lambda_{2}^{2}+36\lambda_{2}^{4}.
\end{align*}
We must choose some values for $\lambda_{1}$ and $\lambda_{2}$ to study the behavior of the trajectories, otherwise, it is so difficult. For example, by taking $\lambda_{1}=0.1759$ and $\lambda_{2}=i$, CP7 is an attractor with $\omega_{\mathrm{eff.}}=-0.99$.
$\blacktriangle$

$\blacktriangledown$\; \textbf{CP8:}\\
This point has the same status as the previous point in the sense that one of $\lambda_{1}$ or $\lambda_{2}$ must be a purely imaginary number. Let us again suppose $\lambda_{2}=ci$ where $c$ is a real positive number.\\
The following exact solution after the pullback of CP8 is obtained:
\begin{align*}
&\Theta =\frac{2\lambda_{1}^{2}\left( \lambda_{2}^{2}-1 \right)+2\lambda_{2}^{2}}{\lambda_{1}^{2}\lambda_{2}^{2}t},
\quad \sigma =\frac{-1}{\sqrt{3}}\frac{\lambda_{1}^{2}\left( \lambda_{2}^{2}+2 \right)-2\lambda_{2}^{2}}{\lambda_{1}^{2}\lambda_{2}^{2}t},\\
&\psi_{1}=\frac{2}{\lambda_{1}t}, \quad \psi_{2}=\frac{2}{\lambda_{2}t},\\
&V_{1}=\frac{2\lambda_{1}^{2}\left( \lambda_{2}^{2}-2 \right)+4\lambda_{2}^{2}}{\lambda_{1}^{4}\lambda_{2}^{2}t^{2}},
\quad V_{2}=\frac{2\lambda_{1}^{2}\left(2 -\lambda_{2}^{2} \right)-4\lambda_{2}^{2}}{\lambda_{1}^{2}\lambda_{2}^{4}t^{2}},\\
&\varphi_{1}=\frac{2\ln (t)}{\lambda_{1}}+c_{1},
\quad \varphi_{2}=\frac{2\ln (t)}{\lambda_{2}}+c_{2},
\end{align*}
where $c_{1}$ and $c_{2}$ are constants of integration.
Therefore, the consequence of $\lambda_{2}$ being purely imaginary number is that the scalar field $\varphi_{2}$ becomes purely imaginary function.\\
The corresponding eigenvalues are:
\begin{align*}
&\varepsilon_{1}=\frac{\lambda_{1}^{2}\left( 2-\lambda_{2}^{2} \right)-2\lambda_{2}^{2}}{2\lambda_{1}^{2}\left(\lambda_{2}^{2}-1 \right)+2\lambda_{2}^{2}},\\
&\varepsilon_{2,3}=\frac{2\lambda_{1}^{2}-2\lambda_{2}^{2}
-\lambda_{1}^{2}\lambda_{2}^{2}\pm \sqrt{B_{1}}}{4\lambda_{1}^{2}\left(  \lambda_{2}^{2}-1\right)+4\lambda_{2}^{2}},\\
&\varepsilon_{4,5}=\frac{2\lambda_{1}^{2}-2\lambda_{2}^{2}
-\lambda_{1}^{2}\lambda_{2}^{2}\pm \sqrt{B_{2}}}{4\lambda_{1}^{2}\left(  \lambda_{2}^{2}-1\right)+4\lambda_{2}^{2}},
\end{align*}
where
\begin{align*}
B_{1}&=\sqrt{\lambda_{1}^{4}\left( -7\lambda_{2}^{4}-4\lambda_{2}^{2}+36 \right)+\lambda_{1}^{2}\left( 4\lambda_{2}^{4}-72\lambda_{2}^{2} \right)
+36\lambda_{2}^{4}},\\
B_{2}&=\sqrt{\lambda_{1}^{4}\left( -7\lambda_{2}^{4}+12\lambda_{2}^{2}+4 \right)-4\lambda_{1}^{2}\left( 3\lambda_{2}^{4}+2\lambda_{2}^{2} \right)
+4\lambda_{2}^{4}},
\end{align*}
Three types of universe can be achieved as follows:\\
1- Under the conditions $\{c>\sqrt{2}, \; \& \; 0<\lambda_{1}<\sqrt{2}c/\sqrt{c^{2}-2} \}$ and also $\{ \lambda_{1}>0,\; \& \; 0<c \leq \sqrt{2} \}$, CP8 is a saddle point in B-III universe. Isotropization does not happen for it because the ratio $\sigma / \Theta$ remains constant. But inflation is achieved as power-law type ($a_{\mathrm{ave.}} \sim t^{\left[2\lambda_{1}^{2}\left( \lambda_{2}^{2}-1\right)
+2\lambda_{2}^{2}\right]/(3\lambda_{1}^{2}\lambda_{2}^{2})}$) with the EoS parameter in the range $-1<\omega_{\mathrm{eff.}} < -1/3$ which is accelerating era. This model solves the horizon and flatness problems because the scale factor grows faster than the horizon.\\
For limiting ranges namely $\lambda_{1}=0$ or $\lambda_{2}=0$, there is the same status with the difference that in these cases we have exponential inflation instead of power-law.\\
2- For $\lambda_{1}=\sqrt{2}c/\sqrt{c^{2}-2}$, we have a saddle CP in B-I universe which for it again isotropy can be reached without inflation.
The amount of EoS parameter, $\omega_{\mathrm{eff.}}=-1/3$, confirms that inflation cannot occur.\\
3- KS universe contains CP8 under the condition $\{c>\sqrt{2},\; \& \; \lambda_{1}>\sqrt{2}c/\sqrt{c^{2}-2}\}$. In this case, the nature of CP8 is attractor with $\omega_{\mathrm{eff.}}>-1/3$. Like previous KS cases, neither inflation nor isotropization occurs at it.
$\blacktriangle$

$\blacktriangledown$\; \textbf{CP9:}\\
This point is similar to CP6 of class-1. CP9 identifies a family of critical points so that the coordinates $(S, P_{1}, P_{2})$ form a hyperboloid of one sheet.\\
The pull-back of points out of CP9 gives an exact solution of the form
\begin{align*}
&\Theta = \frac{1}{t}, \quad \sigma = \frac{S_{0}}{t}, \quad \psi_{1}=\frac{P_{10}}{t},
\quad \psi_{2}=\frac{P_{20}}{t}, \quad V_{1}=V_{2}=0,\\
&\varphi_{1}(t)=P_{10} \ln (t)+c_{1},
\quad \varphi_{2}(t)=P_{20} \ln (t)+c_{2},
\end{align*}
where $c_{1}$ and $c_{2}$ are constants of integration.\\
All the points of this family belong to the B-I universe and
never inflation can occur at it because $a_{\mathrm{ave.}} \sim t^{1/3}$. The scale factor expands at a slower rate than the horizon size and so the horizon and flatness problems are not solved. But the occurrence of isotropization is feasible when $S_{0}=0$.\\
For the linearization of system around a given point $(S_{0}, 0, 0, P_{10}, P_{20})$ we obtain the following eigenvalues:
\begin{align*}
&\varepsilon_{1}=1-\frac{\lambda_{1}}{2},\quad
\varepsilon_{2}=1-\frac{\lambda_{2}}{2},\\
&\varepsilon_{3}=\frac{\sqrt{18\left( 2-3P_{10}^{2}+3P_{20}^{2} \right)}}{9}
+\frac{4}{3}=\frac{2}{\sqrt{3}}S_{0}+\frac{4}{3},
\\
&\varepsilon_{4}=\varepsilon_{5}=0.
\end{align*}
As is seen, this system has at least 2D center manifold. To demonstrate what the whole is like, we choose two sample points as $(P_{10},P_{20})=(1,1)$ (without isotropization) and $(P_{10},P_{20})=(1,1/\sqrt{3})$ (with isotropization).\\
$\maltese$ \textbf{The case $(P_{10},P_{20})=(1,1)$ (without isotropization):}\\
If the system is transformed as $(S,U_{1},U_{2},P_{1},P_{2})|_{\mathrm{CP9}} \to (s,u_{1},u_{2},p_{1},p_{2})=(0,0,0,0,0)$ and then we use the following
transformations
\begin{align*}
&s=\frac{\sqrt{3}}{2}(z+E-R), \quad u_{1}=x, \quad u_{2}=y,\\
&p_{1}=z+R, \quad p_{2}=z+E,
\end{align*}
we arrive at the following topologically equivalent system
\begin{align*}
x^{\prime}=&\left( 1-\frac{\lambda_{1}}{2} \right)x,\\
y^{\prime}=&\left( 1-\frac{\lambda_{2}}{2} \right)y,\\
z^{\prime}=&2z,\\
E^{\prime}=&\frac{-1}{4}E^{4}+\frac{5}{4}ER^{3}+\frac{7}{4}RE^{3}
-\frac{11}{4}E^{2}R^{2}+\frac{3}{32}E^{5}-\frac{9}{8}RE^{4}\\&
+\frac{69}{16}E^{3}R^{2}-\frac{45}{8}R^{3}E^{2}+\frac{75}{32}ER^{4},\\
R^{\prime}=&\frac{5}{4}R^{4}-\frac{1}{4}RE^{3}-\frac{11}{4}ER^{3}
+\frac{7}{4}E^{2}R^{2}+\frac{75}{32}R^{5}-\frac{45}{8}ER^{4}\\
&+\frac{69}{16}E^{2}R^{3}-\frac{9}{8}R^{2}E^{3}+\frac{3}{32}RE^{4},
\end{align*}
where we have exploited the following mappings:
\begin{align*}
h_{1}(E,R)=&c_{1}E^{2}+c_{2}ER+c_{3}R^{2},\\
h_{2}(E,R)=&c_{4}E^{2}+c_{5}ER+c_{6}R^{2},\\
h_{2}(E,R)=&c_{7}E^{2}+c_{8}ER+c_{9}R^{2}.
\end{align*}
\begin{figure}[ht]
\centering
\includegraphics[width=10cm,height=10cm]{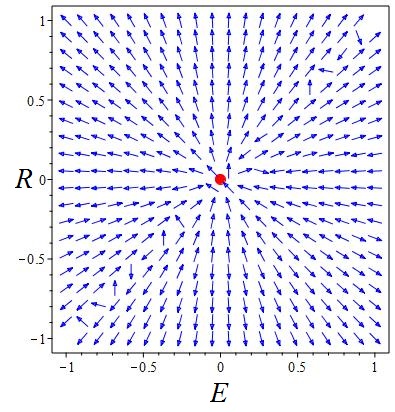}\\
\caption{This figure demonstrates phase portrait of 2d center manifold of CP9 for the case $(P_{10},P_{20})=(1,1)$.}\label{fig3}
\end{figure}
\begin{figure}[ht]
\centering
\includegraphics[width=10cm,height=10cm]{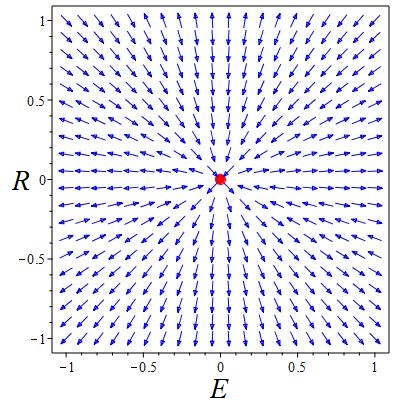}\\
\caption{This figure indicates phase portrait of 2d center manifold of CP9 for the case $(P_{10},P_{20})=(1,1/\sqrt{3})$.}\label{fig4}
\end{figure}
Therefore, the system at CP9 has saddle nature. The behavior of 2d center manifold has been demonstrated in fig.~\ref{fig3}.

$\maltese$ \textbf{The case $(P_{10},P_{20})=(1,1/\sqrt{3})$ (with isotropization):}\\
Performing a transformation as $(S,U_{1},U_{2},P_{1},P_{2})|_{\mathrm{CP9}} \to (s,u_{1},u_{2},p_{1},p_{2})=(0,0,0,0,0)$ and then applying the transformation
\begin{align*}
&s=\frac{1}{2}z+R, \quad u_{1}=x, \quad u_{2}=y,\\
&p_{1}=\sqrt{3}z+\frac{1}{\sqrt{3}}E, \quad p_{2}=z+E,
\end{align*}
our system will be expressed in a new coordinate $(x,y,z,E,R)$. By the use of three mappings
\begin{align*}
h_{1}(E,R)=&c_{1}E^{2}+c_{2}ER+c_{3}R^{2},\\
h_{2}(E,R)=&c_{4}E^{2}+c_{5}ER+c_{6}R^{2},\\
h_{2}(E,R)=&c_{7}E^{2}+c_{8}ER+c_{9}R^{2}.
\end{align*}
we arrive at the following system which is topologically equivalent to the original system:
\begin{align*}
&x^{\prime}=\left( 1-\frac{\lambda_{1}}{2}\right)x,\\
&y^{\prime}=\left( 1-\frac{\lambda_{2}}{2\sqrt{3}}\right)y,\\
&z^{\prime}=\frac{4}{3}z,\\
&E^{\prime}=-\sqrt{3}\;ER^3+\frac{1}{\sqrt{3}}RE^3+\frac{15}{8}ER^4
-\frac{5}{4}E^{3}R^{2}+\frac{5}{24}E^{5},\\
&R^{\prime}=-\sqrt{3}\;R^{4}+\frac{1}{\sqrt{3}}E^{2}R^{2}+\frac{5}{24}RE^{4}
-\frac{5}{4}E^{2}R^{3}+\frac{15}{8}R^{5}.
\end{align*}
As is clear from fig.~\ref{fig4}, this 2d center manifold is also saddle at the origin and thus is a physical point.
$\blacktriangle$

\section{Conclusion}
In this paper, we investigated inflation and isotropization in the quintom model in the Bianchi-I, Bianchi-III, and Kantowski-Sachs backgrounds. First, we studied inherent properties and generalized Heusler's proposition.
Since in this proposition, there was no definitive answer to the question of isotropization for an exponential potential in either a B-I or a B-III model because $\Theta$ and $\sigma$ vanished together, we scrutinized our system utilizing dynamical systems in multiplicative and collective modes of exponential potentials. The outcomes of this study are as follows:
\begin{itemize}
	 \item For Kantowski-Sachs background, we found that neither inflation nor isotropization occurred in both modes.
  \item For the Bianchi-I universe in the multiplicative mode of potentials, isotropy can be reached without inflation which is consistent with Collins and Hawking’s result. However, we found a case that neither inflation nor isotropization occurred. For the collective mode, the status was so different: In one case, both inflation and isotropization happened. For another case, the universe isotopized without any condition while inflation occurred when one of the scalar fields was a purely imaginary function. It is worth noting that the corresponding potential remained a real positive function. Besides, there were cases in which isotropization happened without inflation, and there were also cases in which neither inflation nor isotropization occurred.
  \item For the Bianchi-III case, in multiplicative mode, inflation was reached without isotropization. For collective mode, in one case there was the same status. But, there were some cases in which the occurrence of inflation was under the condition that one of the scalar fields had to be a purely imaginary function. However, again, the corresponding potential remained a real scalar field. Moreover, there was one common case in which neither inflation nor isotropization occurred.
\end{itemize}
Therefore we conclude that the result of Collins and Hawking which demonstrates that ordinary matter can achieve isotropy without inflation within the Bianchi-I universe is true just for multiplicative mode. Furthermore, Burd and Barrow concluded that if inflation occurs then isotropy is always reached~\cite{BBa}. This statement did not come true in any mode. Hence, it seems that their conclusion is true just for a single scalar field.

\end{document}